\documentclass{aa}  

\usepackage{graphicx}
\usepackage{txfonts}
\usepackage{natbib}
\bibpunct{(}{)}{;}{a}{}{,} 

\begin{document}

\title{Benchmarking Gaia DR3 Apsis with the Hyades and Pleiades open clusters}

\author{Wolfgang Brandner\inst{1}
    \and
    Per Calissendorff\inst{2}
    \and
    Taisiya Kopytova\inst{1}
    }

\institute{Max-Planck-Institut f\"ur Astronomie, K\"onigstuhl 17, 69117 Heidelberg, Germany\\                 \email{brandner@mpia.de}
    \and
    Department of Astronomy, University of Michigan, Ann Arbor, MI 48109, USA
    }
    
\date{Received 2 May 2023 / Accepted 5 June 2023}

\abstract
{The Gaia astrophysical parameters inference system (Apsis) provides astrophysical parameter estimates for several to hundreds of millions of stars.}
{We aim to benchmark Gaia DR3 Apsis.}
{We compiled approximately 1500 bona fide single stars in the Hyades and Pleiades open clusters for validation of PARSEC isochrones, and for comparison with Apsis estimates. PARSEC stellar isochrones in the Gaia photometric system enable us to assign average ages and metallicities to the clusters, and mass, effective temperature, luminosity, and surface gravity to the individual stars.}
{Apsis does not recover the single-age, single-metallicity characteristic of the cluster populations.  Ages assigned to cluster members seemingly follow the input template for Galactic populations, with earlier-type stars being systematically assigned younger ages than later-type stars. Cluster metallicities are underestimated by 0.1 to 0.2\,dex. Effective temperature estimates are in general reliable. Surface gravity estimates reveal strong systematic errors for specific ranges of the Gaia BP-RP colours.}
{We caution that Gaia DR3 Apsis estimates can be subject to significant systematic uncertainties. Some of the Apsis estimates, such as metallicity, might only be meaningful for statistical studies of the time-averaged Galactic stellar population, but are not recommended to be used for individual stars.}


\maketitle

\section{Introduction} \label{sec:intro}

Precisely determined stellar parameters are key to many astrophysical investigations. The Gaia astrophysical parameters inference system \citep[Apsis,][]{Bailer2013} was devised to extend the scientific content and potential of Gaia beyond its core astrometric mission of providing precise measurements of positions, parallaxes, and proper motions for more than 10$^9$ sources. Gaia Data Release 3 (DR3) includes astrophysical parameters of up to 470\,million sources. The Apsis estimates are based on averaged low-resolution spectra in the Gaia BP and RP bands with a resolution of R=20 to 60 combined with Gaia photometry and astrometry, and on data from the Radial Velocity Spectrograph (RVS), which provides R$\approx$11500 spectra in the wavelength range from 846 to 870\,nm for stars with G$\lesssim$15.2\,mag. This is used to infer stellar astrophysical parameters such as effective temperature, age, mass, surface gravity log\,g, global metallicity [M/H], and elemental abundances \citep{Creevey2022,Fouesneau2022,Delchambre2022}.

Gaia Apsis promises to unlock vastly improved insights into a variety of research topics, such as the solar neighbourhood and young nearby moving groups, stellar structure and evolution, open and globular clusters, and the overall structure of the Milky Way with its distinct streams, bar, and spiral arms as well as its more diffuse disc and halo components. Precise stellar astrophysical parameters are also essential for the study of exoplanets, whose properties are in general determined relative to the properties of the stellar host \citep[e.g.][]{Magrini2022, Berger2023}.

As pointed out by \citet{Bailer2013}, the veracity of the Apsis methods can only be assessed on real data. 
This provides our motivation to use independently determined properties of bona fide single stars in the Hyades and Pleiades open clusters as external validations of, and benchmarks for, Gaia DR3 Apsis estimates.

The structure of the paper is as follows. In Sect.\ \ref{sec:sss}, we review the compilation of the single-star sequence of the Hyades open cluster, and apply the same methodology to the Pleiades to validate PARSEC isochrones. In Sect.\ \ref{sec:GaiaApsis}, we compare cluster ensembles and individual stellar astrophysical properties with Apsis estimates. In Sect.\ \ref{sec:discussion}, we discuss the findings and potential explanations for some of the issues identified. We conclude in Sect.\ \ref{sec:outlook} with recommendations as to the utilisation of DR3 Apsis estimates and an outlook.

\section{Single-star sequences in the Hyades and Pleiades} \label{sec:sss}

\begin{figure}[htb]
\includegraphics[width=0.48\textwidth]{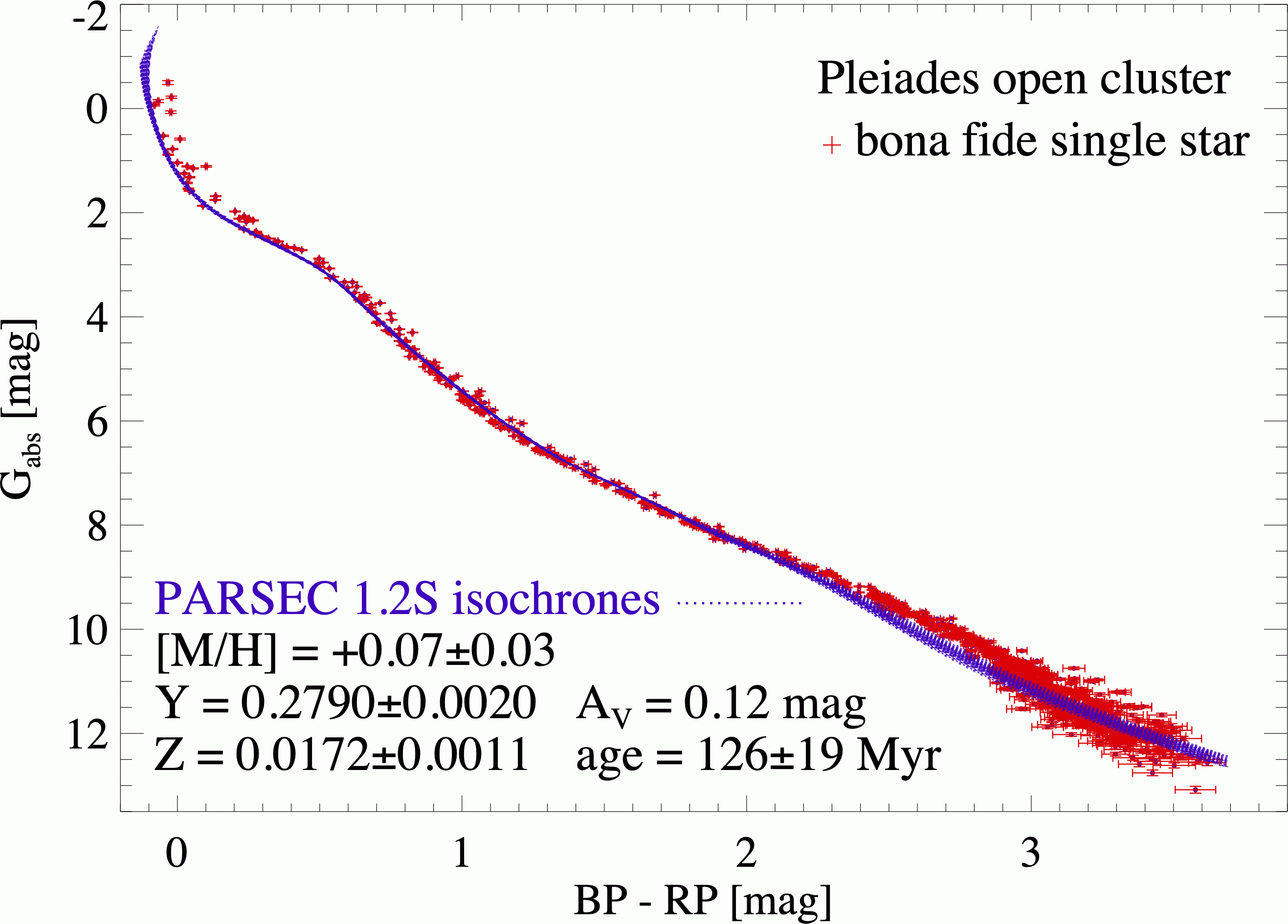}
\caption{Colour--absolute magnitude diagram of bona fide single stars in the Pleiades open cluster based on GAIA DR3. Overplotted is the family of the best-fitting PARSEC isochrones. 
\label{fig:PleiaCMDall}}
\end{figure}

\begin{figure}[htb]
\includegraphics[width=0.48\textwidth]{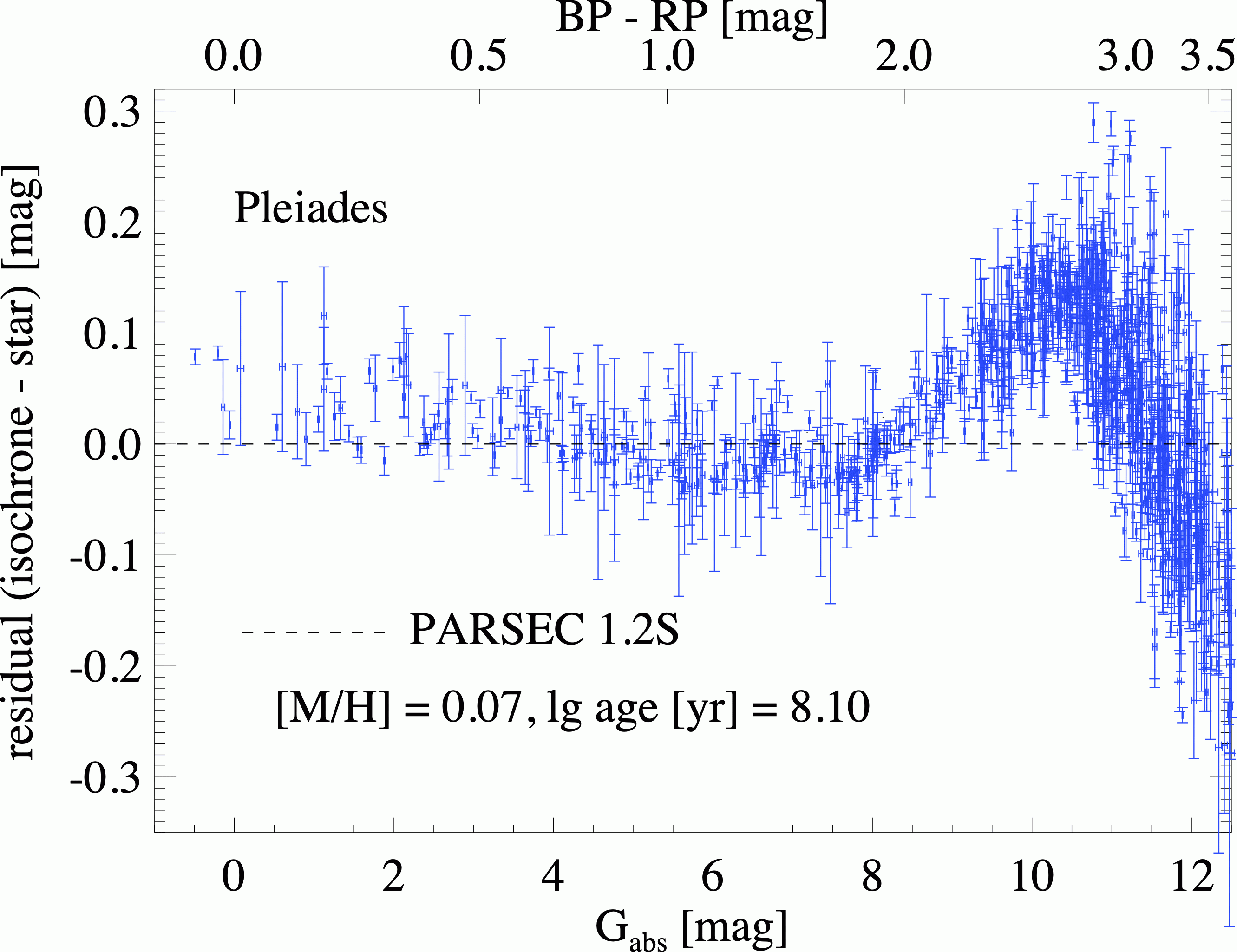}
\caption{Minimum distance in colour--magnitude space of each bona fide single star in the Pleiades from the best-fitting isochrone plotted against the absolute magnitude (lower abscissa) and colour (upper abscissa) on the isochrone. A positive residual corresponds to stars that are brighter (redder) than predicted by the isochrone. A negative residual corresponds to stars that are fainter (bluer) than the isochrone. The uncertainty for each source in G$_{\rm abs}$ corresponds to the Gaia uncertainty in absolute G magnitude, and the uncertainty in the residual corresponds to the quadratically added uncertainties in absolute G magnitude and BP-RP colour. 
\label{fig:IsoResidualsPleiades}}
\end{figure}

\begin{table*}
\caption{Compilation of abundance, $\alpha_{\rm ML}$, and age estimates for the Pleiades open cluster based on isochrone fitting.}             
\label{tab:pleiades_age}      
\centering                          
\begin{tabular}{l l l l l l l l l}        
[Fe/H] & X & Y & Z& $\alpha_{\rm ML}$ & mass range & age  &PD$^1$&reference \\
 &   &   &  & &   [M$_\odot$] & [Myr] & & \\ \hline
 &0.68 & 0.30 & 0.02 &1.6 &[0.8,4.8] & 100&BD&\cite{Meynet1993}\\
  & & & &  &[0.03,4.8]&120&DA& \cite{Bouy2015}$^2$\\ 
  $+0.03\pm0.05$&  &  &   & &[$\approx$2.5]  &$95 \pm 35$ &TY&\cite{Brandt2015}\\
$>+0.29$&  &    & &1.82 &[0.95,4.8]  &$125 \pm 35$ &TY&\cite{Gossage2018}\\
$+0.18${\raisebox{0.5ex}{\tiny$^{+0.29}_{-0.15}$}}&  & & &1.82&[0.6,4.8]  &160{\raisebox{0.5ex}{\tiny$^{+180}_{-60}$}} &2M&\cite{Gossage2018}\\ 
&  &  &  &&[0.2,4.8]  &120$\pm20$&G3&\cite{Heyl2022}\\ \hline
\end{tabular}
    \begin{quote}
        $^1$key to photometric data set (PD): 2M - based on 2MASS photometry \citep{Cutri2003}; BD - based on BDA \citep{Mermilliod1992}; DA - based on {\it DANCe}  \citep{Bouy2013}; G3 - based on {\it GAIA} EDR3 photometry \citep{GAIA_Brown2021}; TY - based on {\it TYCHO} photometry \citep{Hog2000}; $^2$aimed at deriving empirical isochrones, and therefore only a comparison to theoretical isochrones, but no formal isochrone fitting. 
      \end{quote}
\end{table*}

\begin{table*}[htb]
\caption{Astrophysical parameters of bona fide single stars in the Pleiades open cluster \label{tab:Pleia_single_tab}}
\centering 
\resizebox{\textwidth}{!}{%
\begin{tabular}{l l l l l l l l l l l l l}
Gaia DR3 ID & RA &DEC & d & G & mass & $\sigma_{\rm mass}$ & log T$_{\rm eff}$ & $\sigma_{\rm logTeff}$ & log L & $\sigma_{\rm log L}$ & log g &$\sigma_{\rm log g}$  \\ 
& (deg) & (deg) & (pc) & (mag) & (M$_\odot$) & (M$_\odot$) & (K) & (K) & (L$_\odot$) & (L$_\odot$) & (cm/s$^2$) & (cm/s$^2$)\\
    68828225308044288&  52.797083&  24.232163& 129.717&  17.3285& 0.2011&  0.0199&  3.4712&  0.0063& -2.2212&  0.0546&  4.8196&  0.0305 \\
     69335619861034752&  52.816578&  25.255258& 130.680&   8.0884& 1.6837&  0.0389&  3.9026&  0.0076&  0.9233&  0.0421&  4.3043&  0.0040 \\
     69585140280997760&  52.824851&  26.028613& 138.364& 12.3085& 0.7792&  0.0082&  3.6793&  0.0024& -0.6595&  0.0148&  4.6555&  0.0033 \\
     69583177479805824&  52.836332&  25.959207& 144.031& 19.2850& 0.1695&  0.0039&  3.4507&  0.0017& -2.3999&  0.0274&  4.7990&  0.0254 \\
     68051390279853824&  53.002084&  23.774632& 137.339& 13.3061& 0.6515&  0.0073&  3.6234&  0.0032& -0.9824&  0.0175&  4.6816&  0.0045 \\
\end{tabular}
}
\tablefoot{Table 2 is published in its entirety in machine-readable format. The first five entries --- and with some of the columns with uncertainties on the distance estimates and part of the Gaia DR3 photometry  omitted--- are shown here for guidance regarding the form and content of the table.}
\end{table*}

At average distances of $\approx$45\,pc \citep{vanLeeuwen2009,Roeser2011,Gaia_Babusiaux2018} and $\approx$135\,pc \citep[e.g.][]{Percival2005,Heyl2022}, respectively, the Hyades and the Pleiades are among the most nearby open clusters. The vicinity ensures high-quality Gaia astrometric, photometric, and spectroscopic measurements over a wide range of spectral types and stellar masses. Age estimates for the Hyades derived from isochrone fitting are in the range of 500 to 800\,Myr \citep{Brandner2023a}. For the Pleiades, the age estimates derived from isochrone fitting centre around 100 to 125\,Myr (Table \ref{tab:pleiades_age}), which is in good agreement with age estimates based on the lithium-depletion boundary \citep[e.g.][]{Basri1996,Stauffer1998,Dahm2015,Galindo2022}.

The stellar samples of candidate members of the open clusters are based on \citet{GAIA_Smart2021} for the Hyades, and on \citet{Heyl2022} for the Pleiades. In \citet{Brandner2023a}, we processed the Hyades sample by identifying and removing stars with dubious photometry in at least one of the Gaia DR3 G, BP, or RP bands. We then used the Gaia renormalised unit weight error (RUWE) parameter to identify likely astrometric binaries. In a final step, likely photometric (i.e.\ blended) binaries were identified based on their distance from the median stellar sequence in colour--magnitude space, taking into account the uncertainties in absolute brightness and colour. This resulted in a sample of about 600 bona fide single stars in the Hyades open cluster. In \citet{Brandner2023b}, we derived the astrophysical parameters for these stars by comparison with the best-fitting family of isochrones from the PAdova and TRieste Stellar Evolution Code \citep[PARSEC,][]{Bressan2012,Chen2014}. 

Here we apply the same methodology to the Pleiades. The sample of approximately 1300 stars within 10\,pc of the nominal centre of the Pleiades open cluster as defined by \citet{Heyl2022} includes 1191 sources with $-0.2\le$BP-RP$<$3.9\,mag and $-2.0\le$G$_{\rm abs}\le 13.5$\,mag. After rejection of 112 stars with dubious Gaia photometry, and the identification of 171 stars as likely astrometric or photometric binaries, we retain a sample of 908 bona fide single stars. We employ the PARSEC CMD 3.7 web interface\footnote{http://stev.oapd.inaf.it/cmd} to obtain version 1.2S grids of non-rotating isochrones in the Gaia EDR3 photometric system for a range of ages and [M/H], and for a visual extinction of A$_{\rm V} = 0.12$\,mag \citep[][and assuming A$_{\rm V} = 3.1*$E(B-V)]{Meynet1993,vanLeeuwen2009}\footnote{We attribute the higher A$_{\rm V}=0.20$\,mag reported by \citet{Heyl2022} due to their use of strictly solar metallicity [M/H] = 0 isochrones.}. Absolute G$_{\rm abs}$ magnitudes are based on the photo-geometric distance estimates provided by \citet{Bailer2021}. $\chi ^2$ minimisation is used to find the best-fitting isochrones to the 235 brightest (G$_{\rm abs} \le 8.0$\,mag) bona fide single upper-main sequence stars. This yields lg(age [yr]) = $8.10 \pm 0.07$ and [M/H] = $0.07 \pm 0.03$ for the Pleiades open cluster.

The colour--magnitude diagram (CMD) of $\approx$900 bona fide single stars in the Pleiades open cluster, and the family of best-fitting PARSEC isochrones is shown in Fig.\ \ref{fig:PleiaCMDall}. Similar to the Hyades, PARSEC isochrones provide an excellent fit to the observed sequence for G$_{\rm abs}\lessapprox 9.0$\,mag. For 9.0$<$G$_{\rm abs}<$11.0\,mag, the isochrones predict fainter G magnitudes by up to 0.15\,mag and bluer BP-RP colours. For still fainter, mid- to late M-dwarfs, the isochrones tend to predict overly bright G magnitudes and redder BP-RP colours (Fig.\ \ref{fig:IsoResidualsPleiades}).

For each star, we calculate its distances to the best-fitting interpolated isochrone in colour--magnitude space to assign T$_{\rm eff}$, $\log$\,g, $\log$\,L, and (present-day) mass. The uncertainties in astrophysical parameters take into account the uncertainties in the Gaia measurements and also those in the [M/H] and age estimates of the isochrone. The derived properties of the bona fide single stars in the Pleiades are listed in Table \ref{tab:Pleia_single_tab}. The corresponding stellar properties for the Hyades open cluster are presented in Table 1 in \citet{Brandner2023b}.

\section{Comparison with Gaia Apsis} \label{sec:GaiaApsis}

\subsection{Cluster ensemble properties: Age and metallicity}

\begin{figure*}[ht!]
\hbox{
\includegraphics[width=0.48\textwidth]{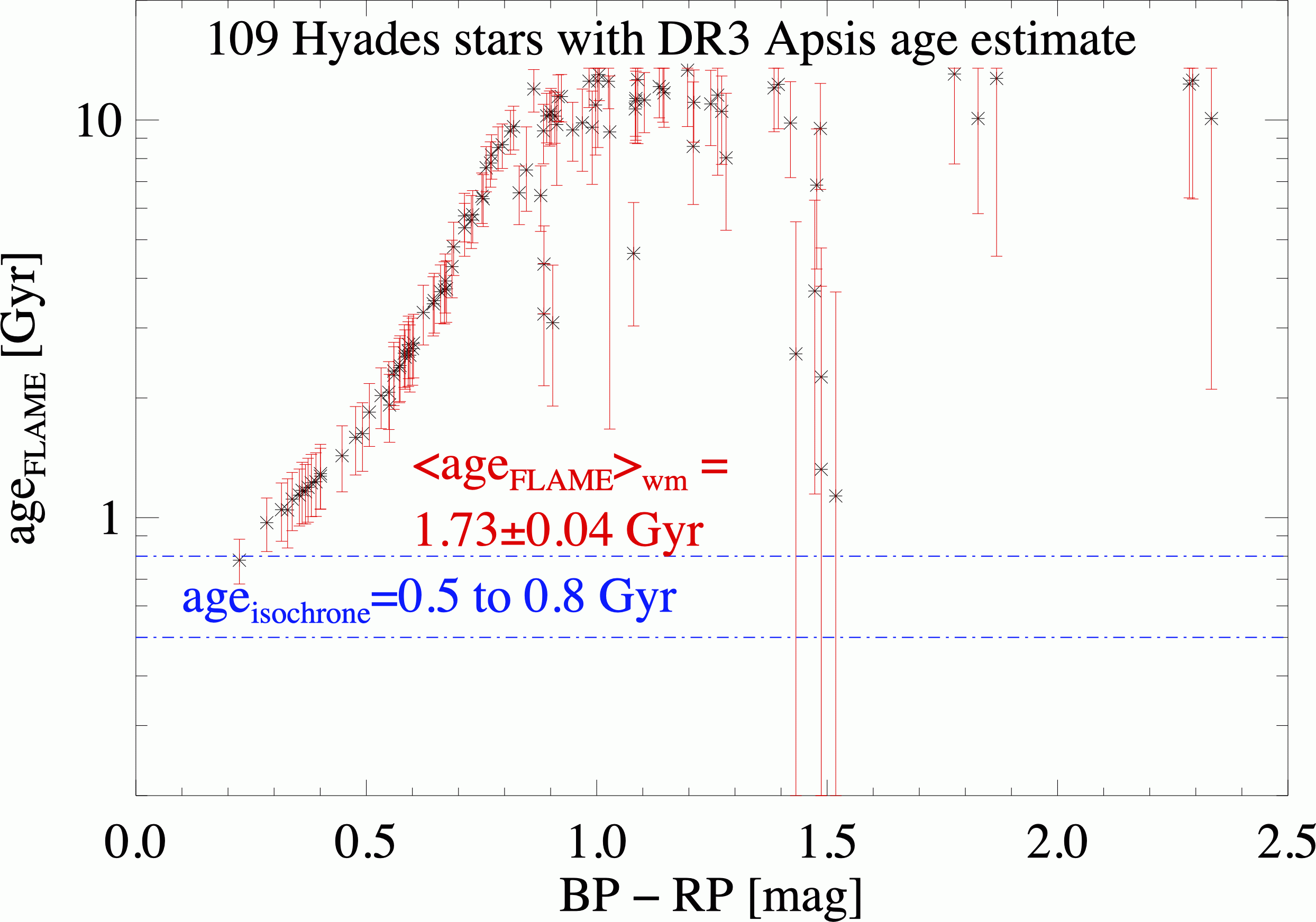}
\includegraphics[width=0.48\textwidth]{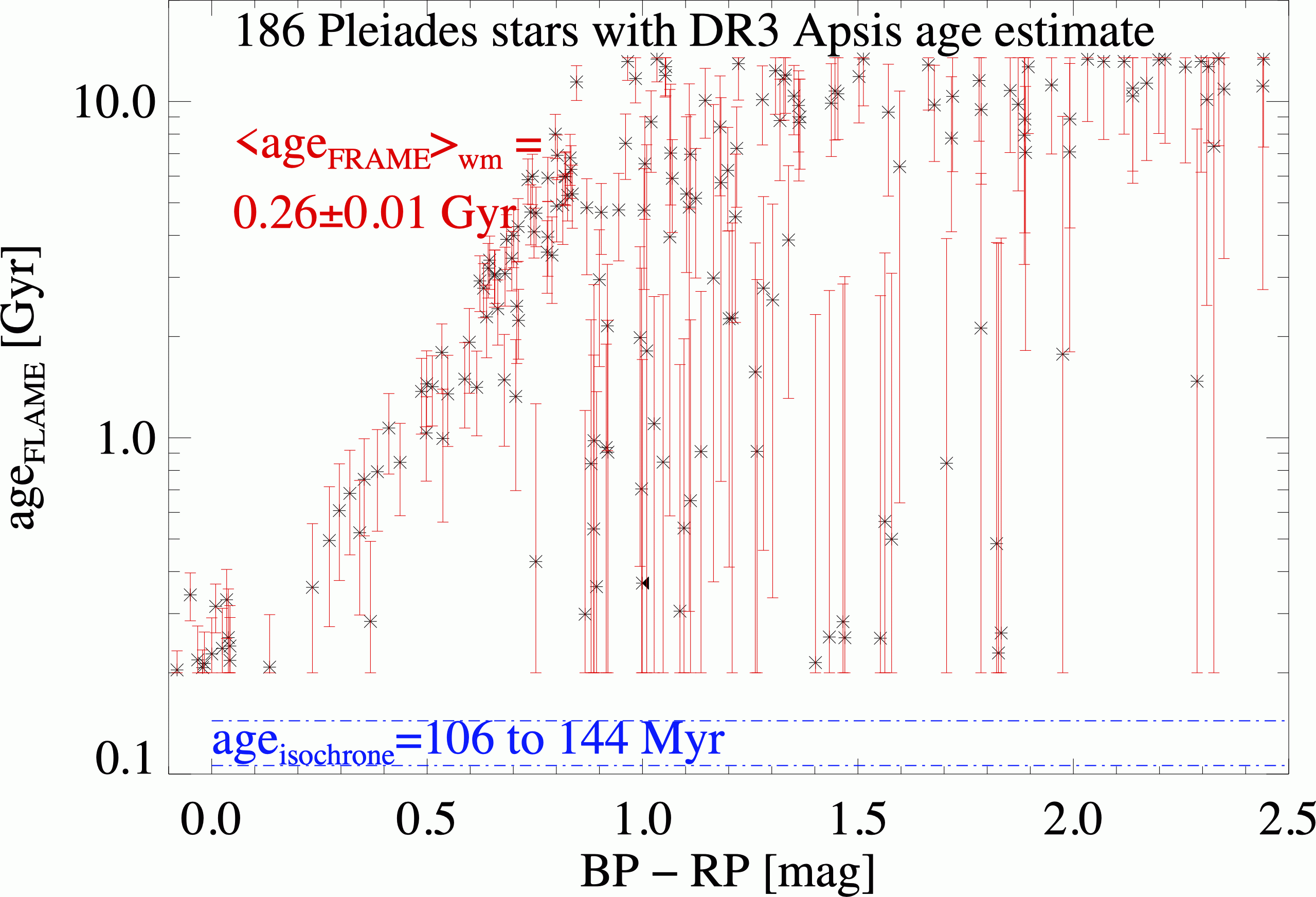}
}
\caption{Gaia DR3 Apsis age estimates as a function of the BP-RP colour for single stars in the Hyades (left) and Pleiades (right). The blue horizontal dashed-dotted lines mark the range of ages assigned to the open clusters by isochrone fitting. For both clusters, Apsis age estimates show a strong correlation with the colour of the star for BP-RP $\le 0.8$\,mag. 
\label{fig:HyaAge}}
\end{figure*}

\begin{figure*}[ht!]
\hbox{
\includegraphics[width=0.48\textwidth]{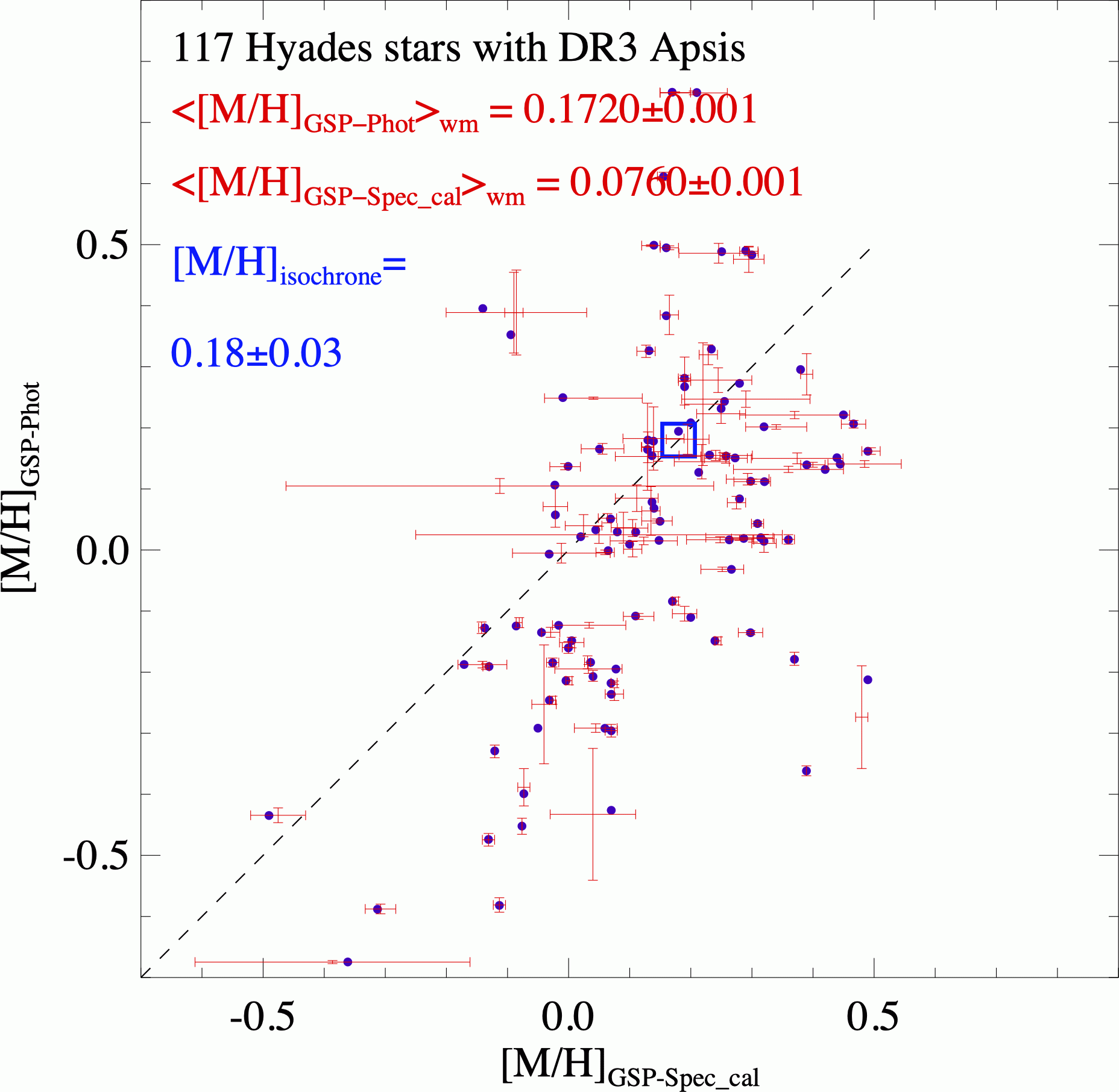}
\includegraphics[width=0.48\textwidth]{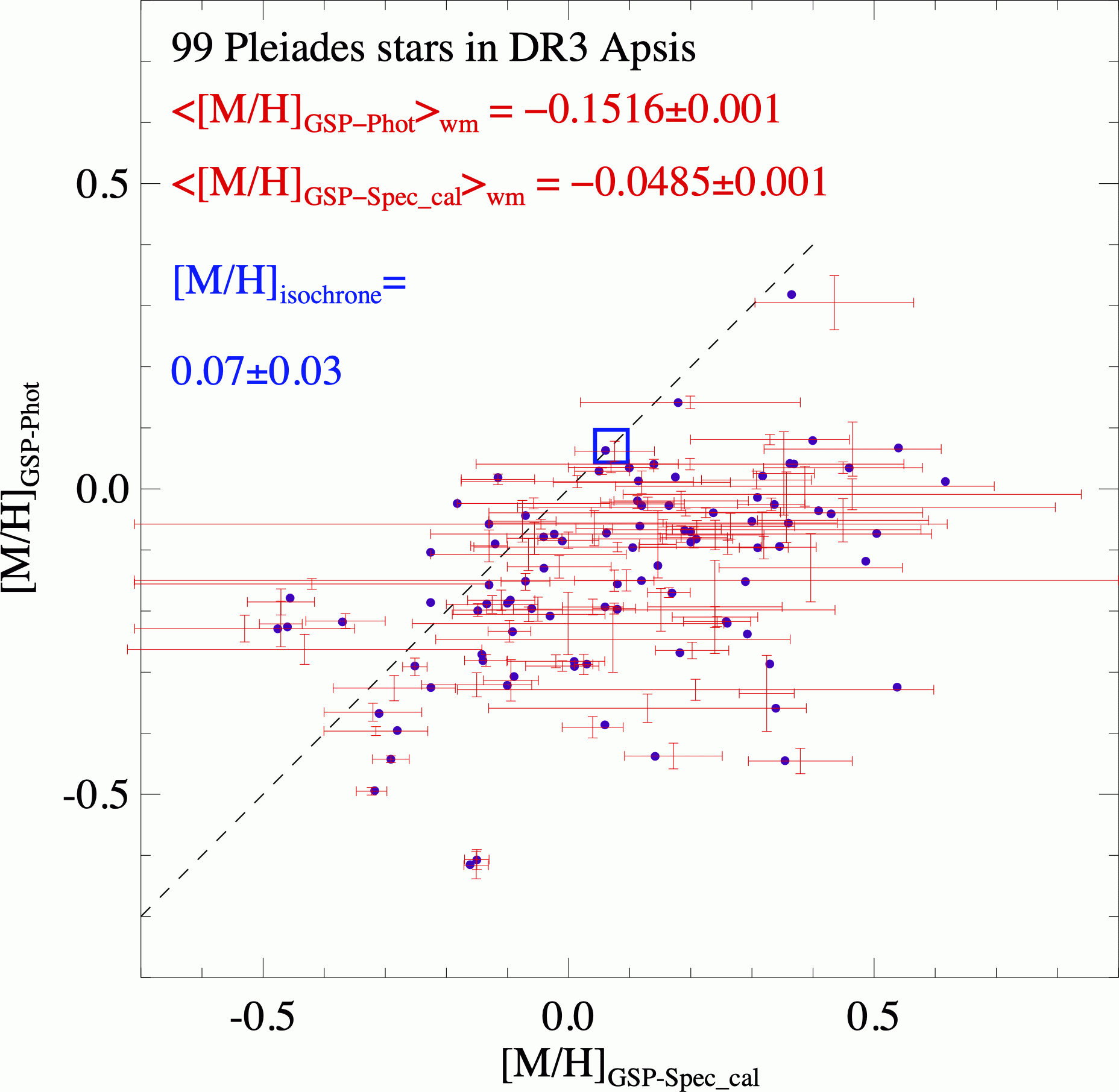}
}
\caption{[M/H] of single stars in the Hyades (left) and Pleiades (right), which have both DR3 Apsis GSP-Phot and recalibrated GSP-Spec estimates. Marked as a black dashed line is the diagonal along which both [M/H] estimates agree. The blue open squares mark the range of [M/H] estimates derived from the best-fitting PARSEC isochrones. Not shown are stars with Apsis estimates of [M/H]$<-0.7$.
\label{fig:HyaMH}}
\end{figure*}

A defining characteristic of stellar populations in open clusters is their small spread in age and abundances \citep{Parmentier2014,Bovy2016,Kopytova2016,Magrini2017,Jeffries2017,Krumholz2020}. In order to test how well Apsis recovers these essential properties, we compare the estimates for individual stars with the ensemble average for each cluster as derived by isochrone fitting. Apsis age estimates originate in the Final Luminosity Age and Mass Estimator (FLAME) module \citep{Bailer2013,Creevey2022}. FLAME in turn relies on input both from the Generalized Stellar Parametrizer – Photometry (GSP-Phot) and the Generalized Stellar Parametrizer – Spectroscopy (GSP-Spec).

Figure\ \ref{fig:HyaAge} shows FLAME mean age estimates along with the lower 16th and 84th quantiles for 109 stars in the Hyades and 186 stars in the Pleiades against their BP-RP colour. FLAME lower and upper age estimates are capped between 0.2 and 13.5\,Gyr\footnote{We note that this contradicts the lower limit of 0.1\,Gyr quoted in the Gaia DR3 documentation release version 1.2 (7.\ Feb.\ 2023), section 11.3.6}. For both clusters, the FLAME age estimates reveal a strong correlation with colour for BP-RP $\le 0.8$\,mag. For BP-RP $> 1.0$\,mag, the majority of age estimates tend to scatter around ages of $\approx$11\,Gyr.
Weighted-mean uncertainties yield $<{\rm age_{FLAME}}>_{\rm wm} = 1.73\pm 0.04$\,Gyr for the Hyades and $<{\rm age_{FLAME}}>_{\rm wm} = 0.26\pm 0.01$\,Gyr for the Pleiades. 

The Apsis GSP-Phot and GSP-Spec modules independently provide [M/H] estimates.
In order to test their consistency, we identified all stars that have [M/H] estimates from both modules. Figure\ \ref{fig:HyaMH} shows the [M/H] estimates along with diagonal dashed lines marking a 1:1 correspondence. GSP-Phot [M/H] are the original published Apsis values. A recalibration using Code Version 1.6 of the gdr3apcal Python package\footnote{\texttt{gdr3apcal} codebase: https://github.com/mpi-astronomy/gdr3apcal} according to \citet{Andrae2022} results in revised [M/H] estimates for Pleiades stars with a wider intrinsic spread; on average these are 1.0\,dex lower than the original Apsis estimates. As a low average [M/H]$\approx -1.1$ appears non-physical for member stars of the Pleiades \citep{Funayama2009,Soderblom2009}, we reject the recalibration of the GSP-Phot [M/H].
GSP-Spec [M/H] estimates are recalibrated according to \citet{RecioBlanco2022}, and using the polynomial coefficients applicable to open clusters. This results in an average correction of $\approx$0.09\,dex towards higher [M/H] values.


\subsection{Individual stellar properties: T$_{\rm eff}$, log\,g, and mass}

\begin{figure*}[ht!]
\vbox{
\hbox{
\includegraphics[width=0.48\textwidth]{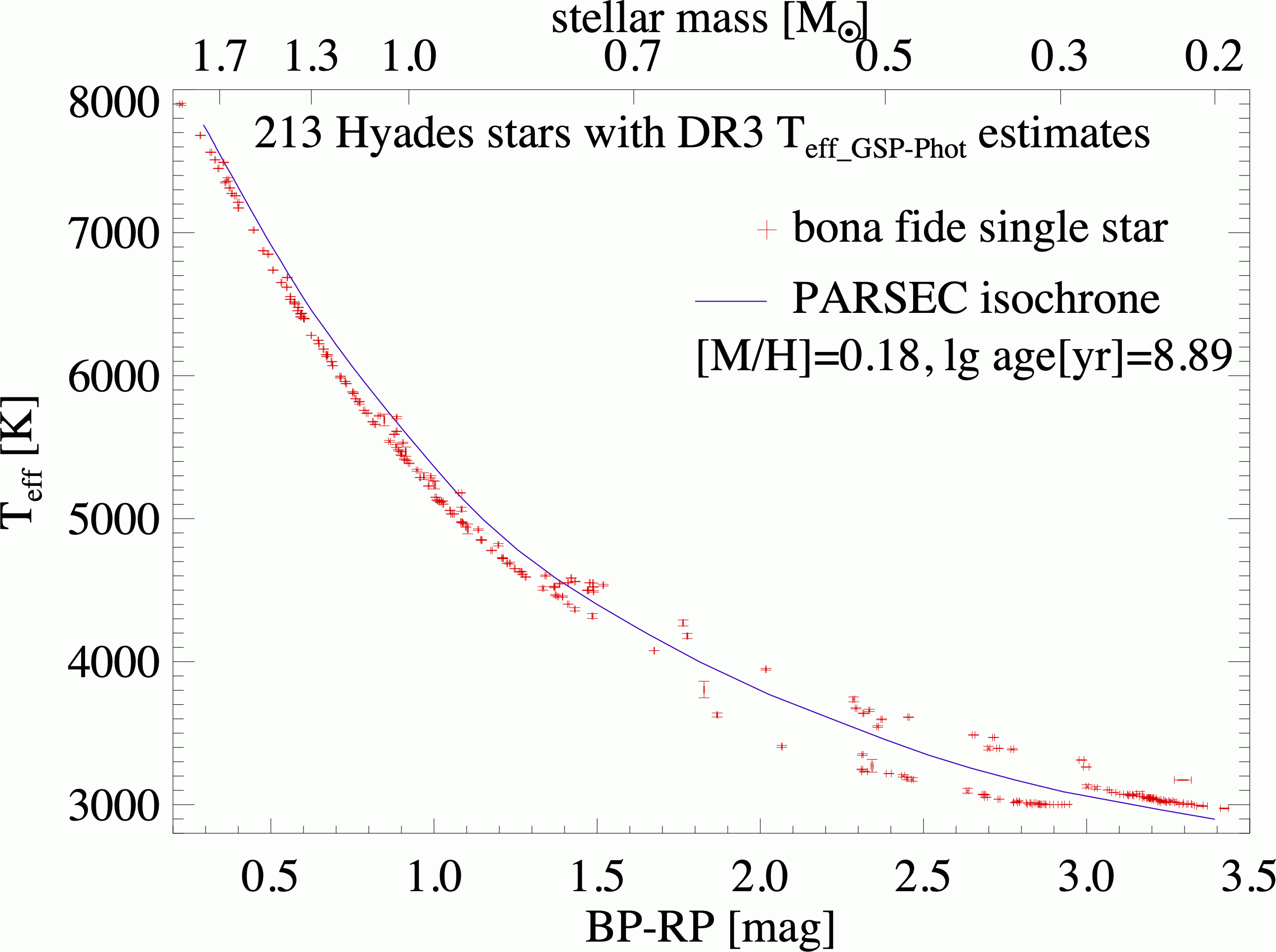}
\includegraphics[width=0.48\textwidth]{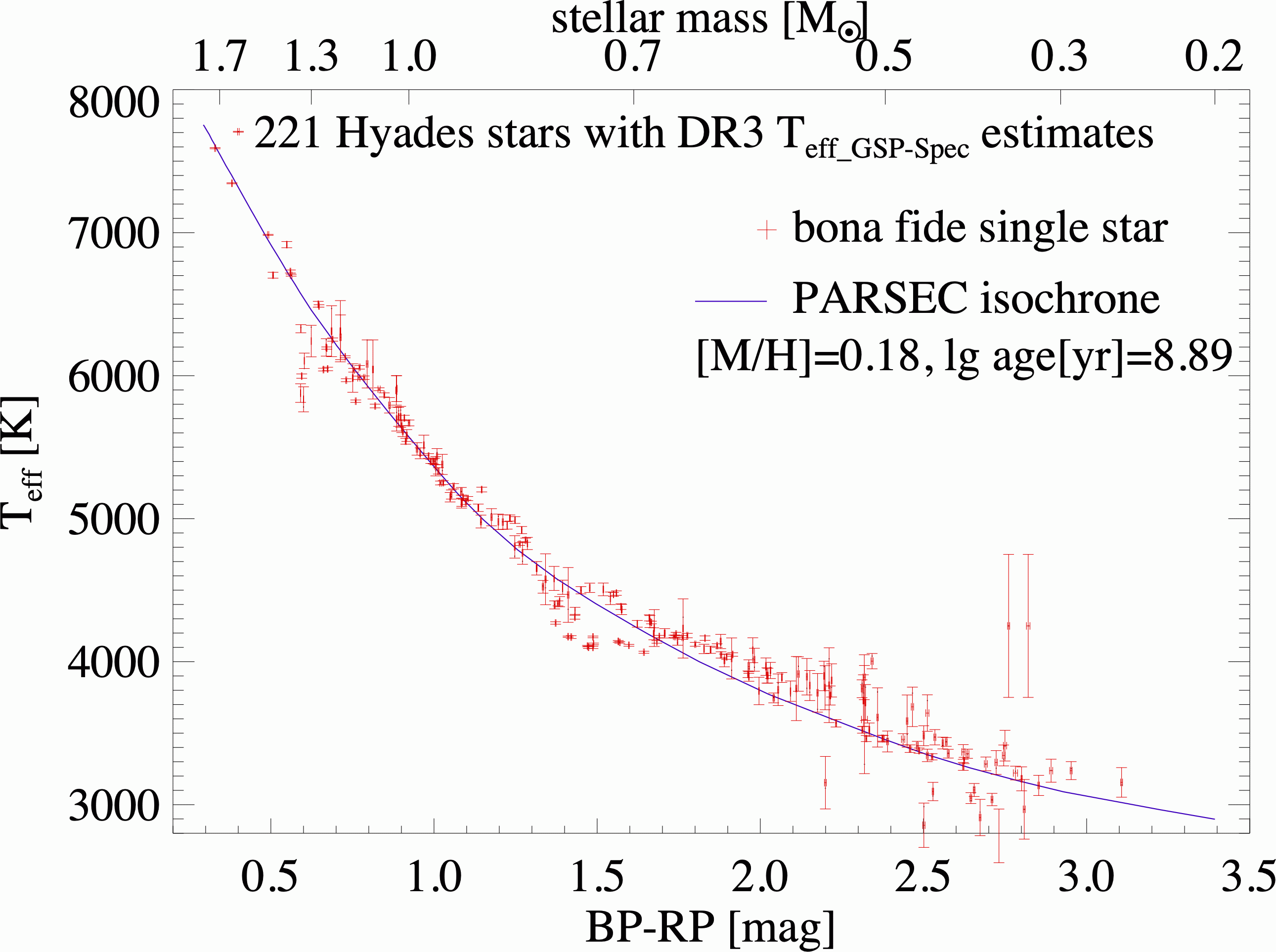}
}
\hbox{
\includegraphics[width=0.48\textwidth]{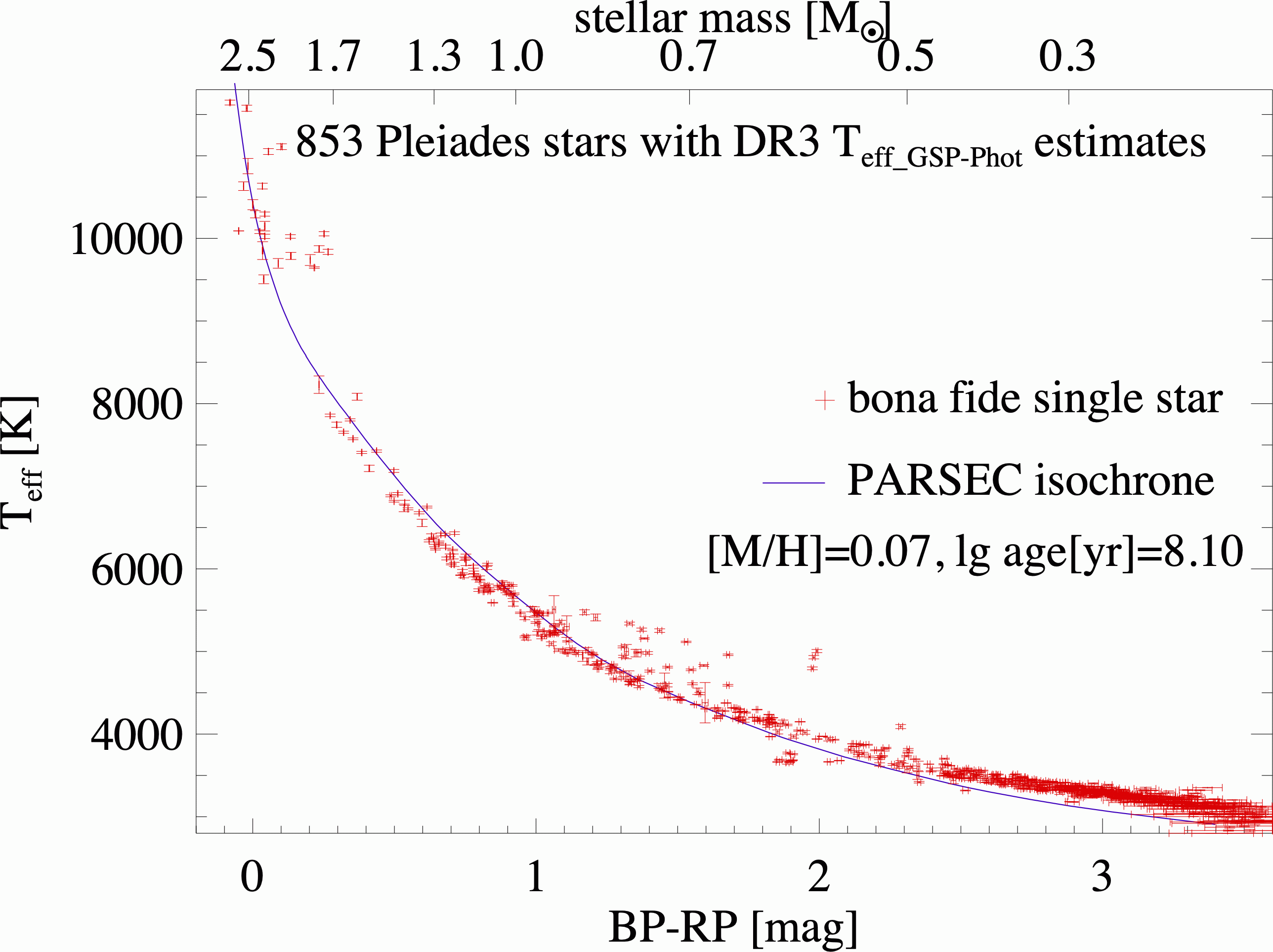}
\includegraphics[width=0.48\textwidth]{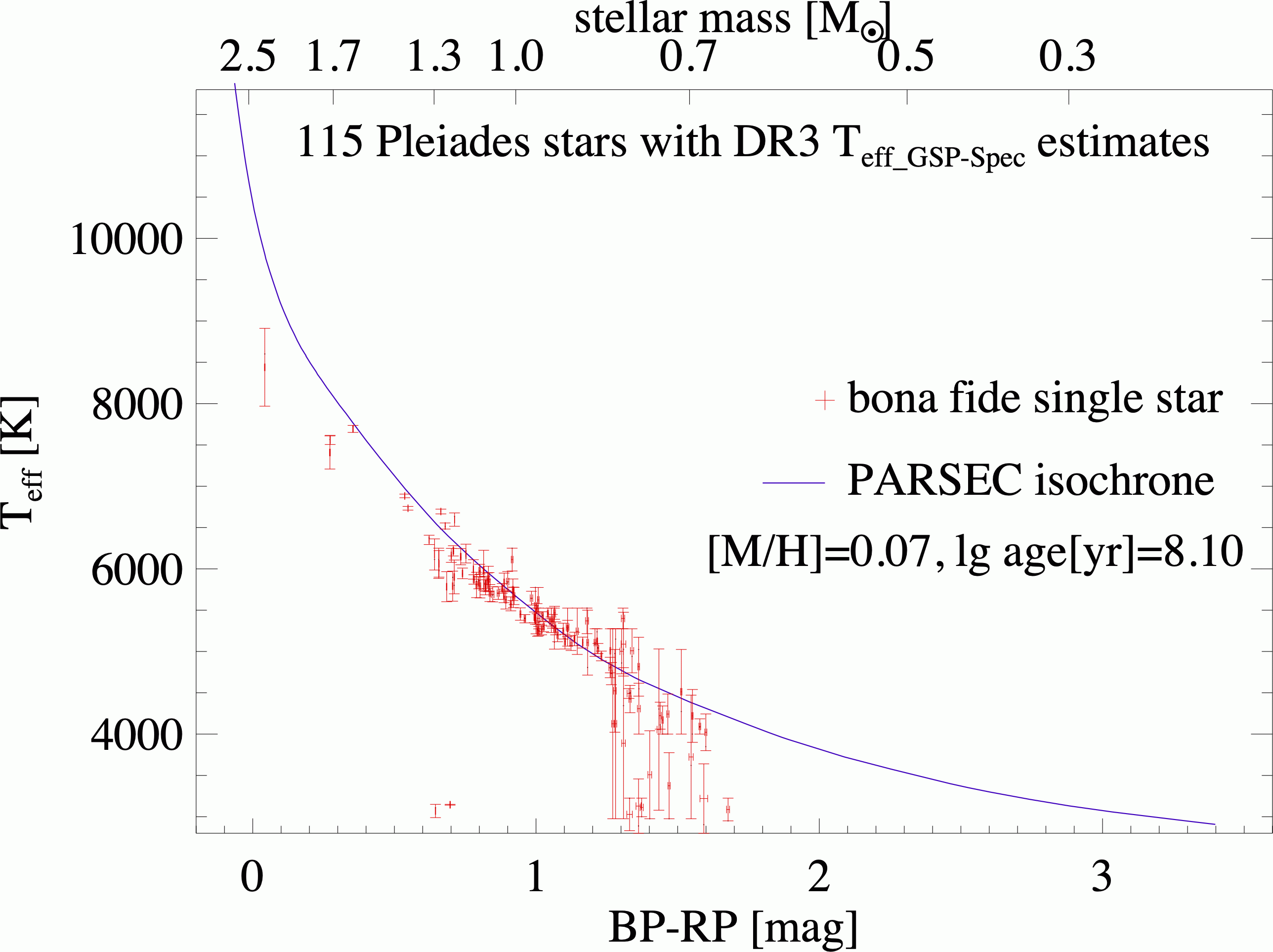}
}
}
\caption{Effective temperature estimates for single stars in the Hyades (top) and Pleiades (bottom) based on GSP-Phot (left) and GSP-Spec (right) as a function of BP-RP colour. Overplotted in blue is the best-fitting PARSEC isochrone.
\label{fig:HyaTeff}}
\end{figure*}

\begin{figure*}[ht!]
\vbox{
\hbox{
\includegraphics[width=0.48\textwidth]{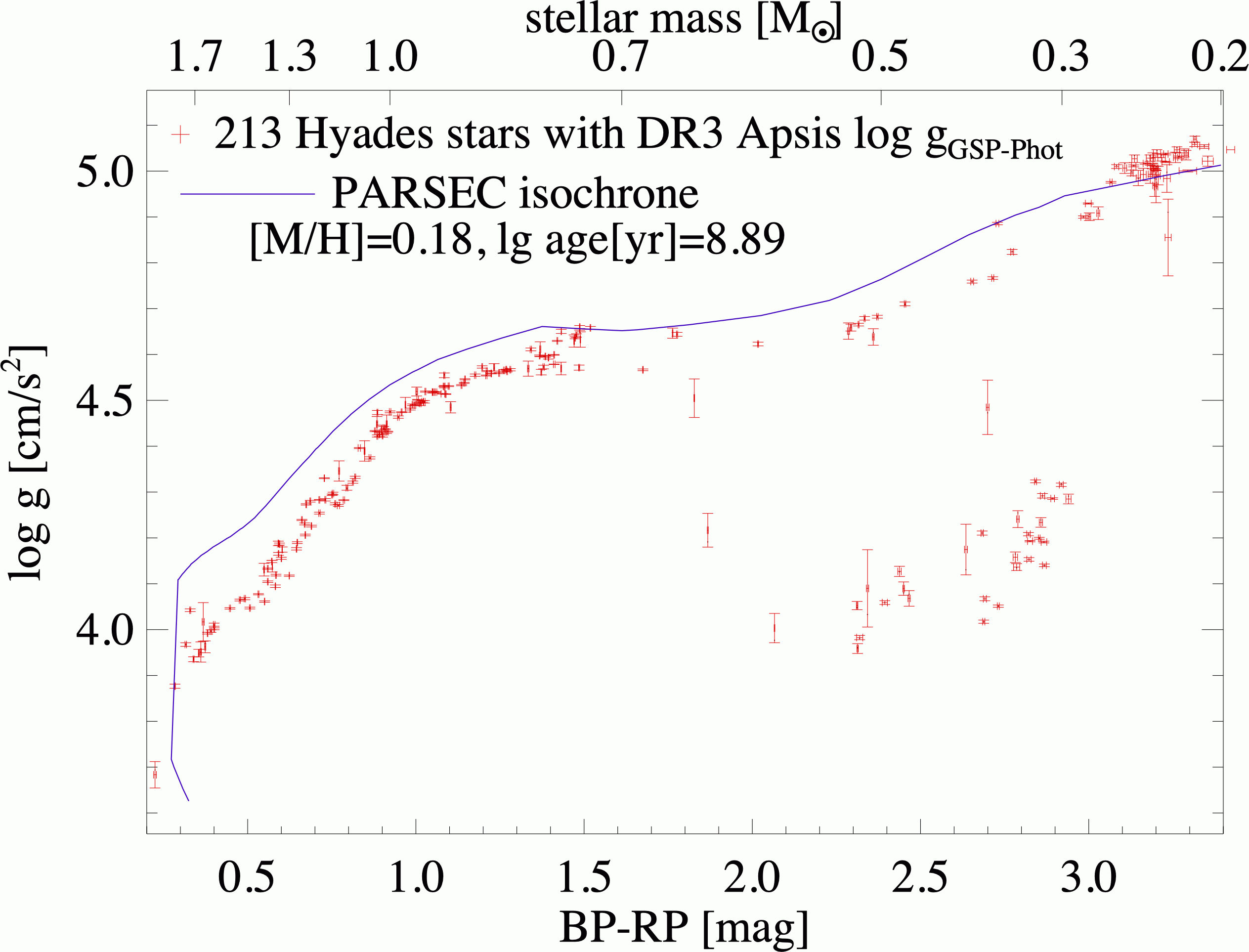}
\includegraphics[width=0.48\textwidth]{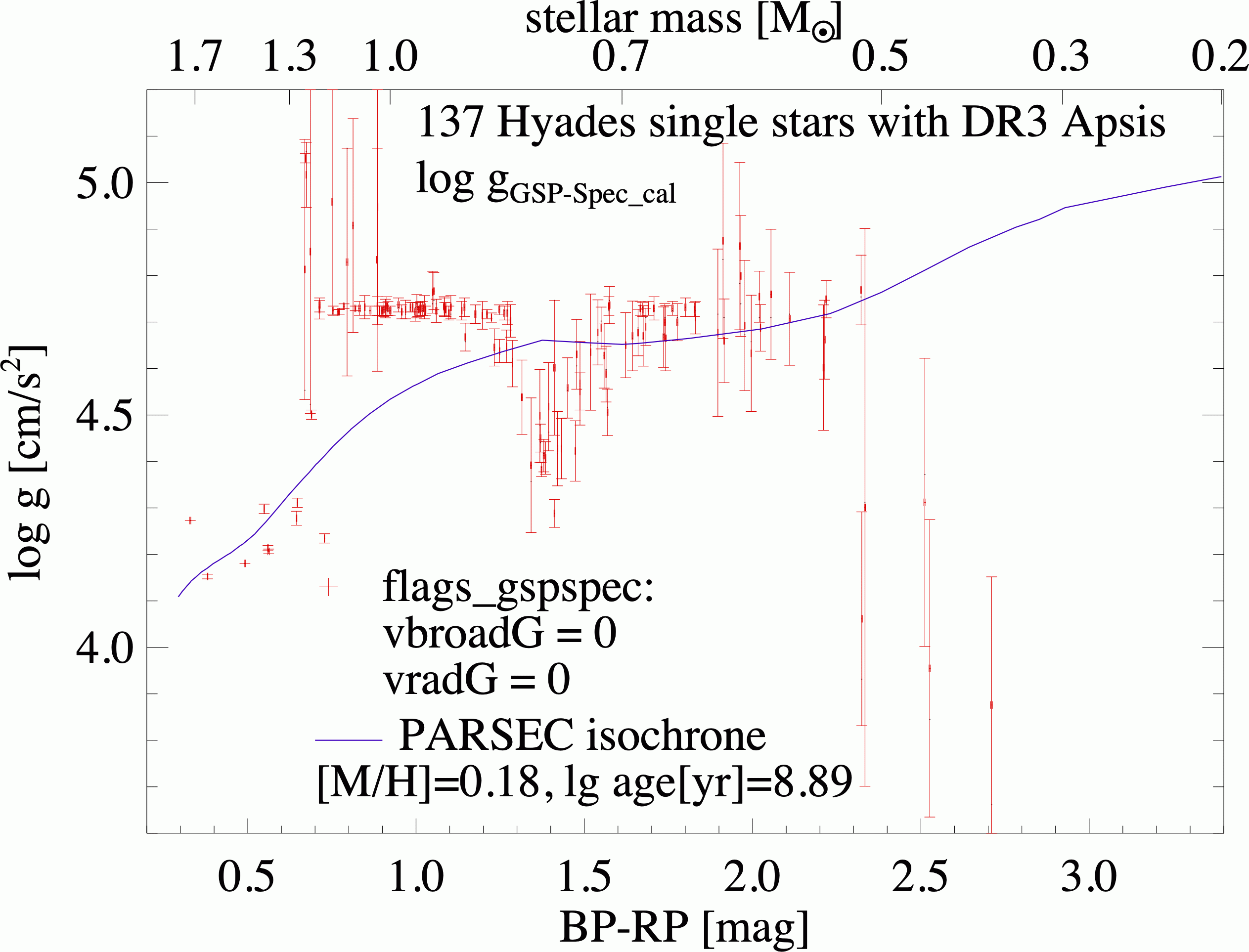}
}
\hbox{
\includegraphics[width=0.48\textwidth]{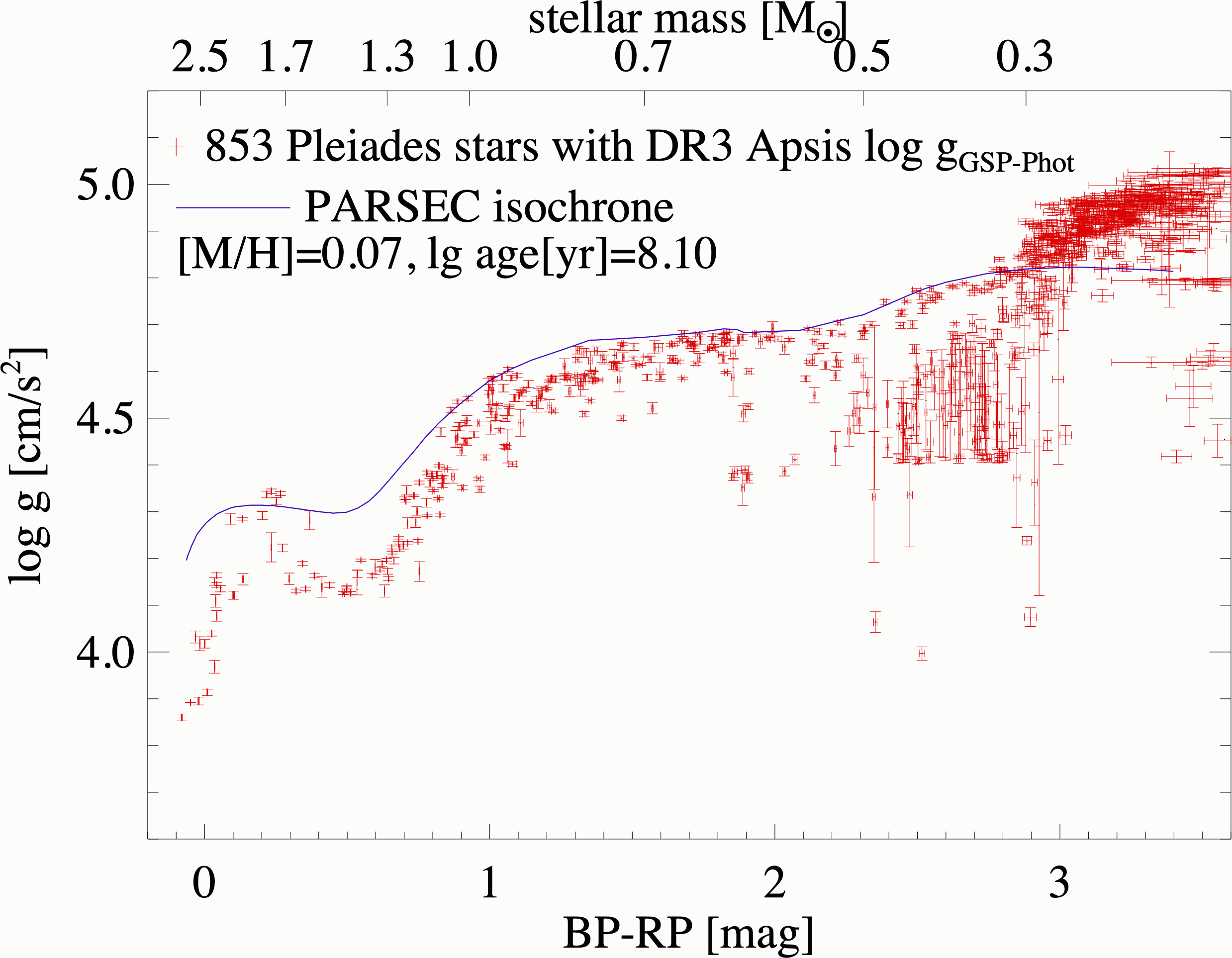}
\includegraphics[width=0.48\textwidth]{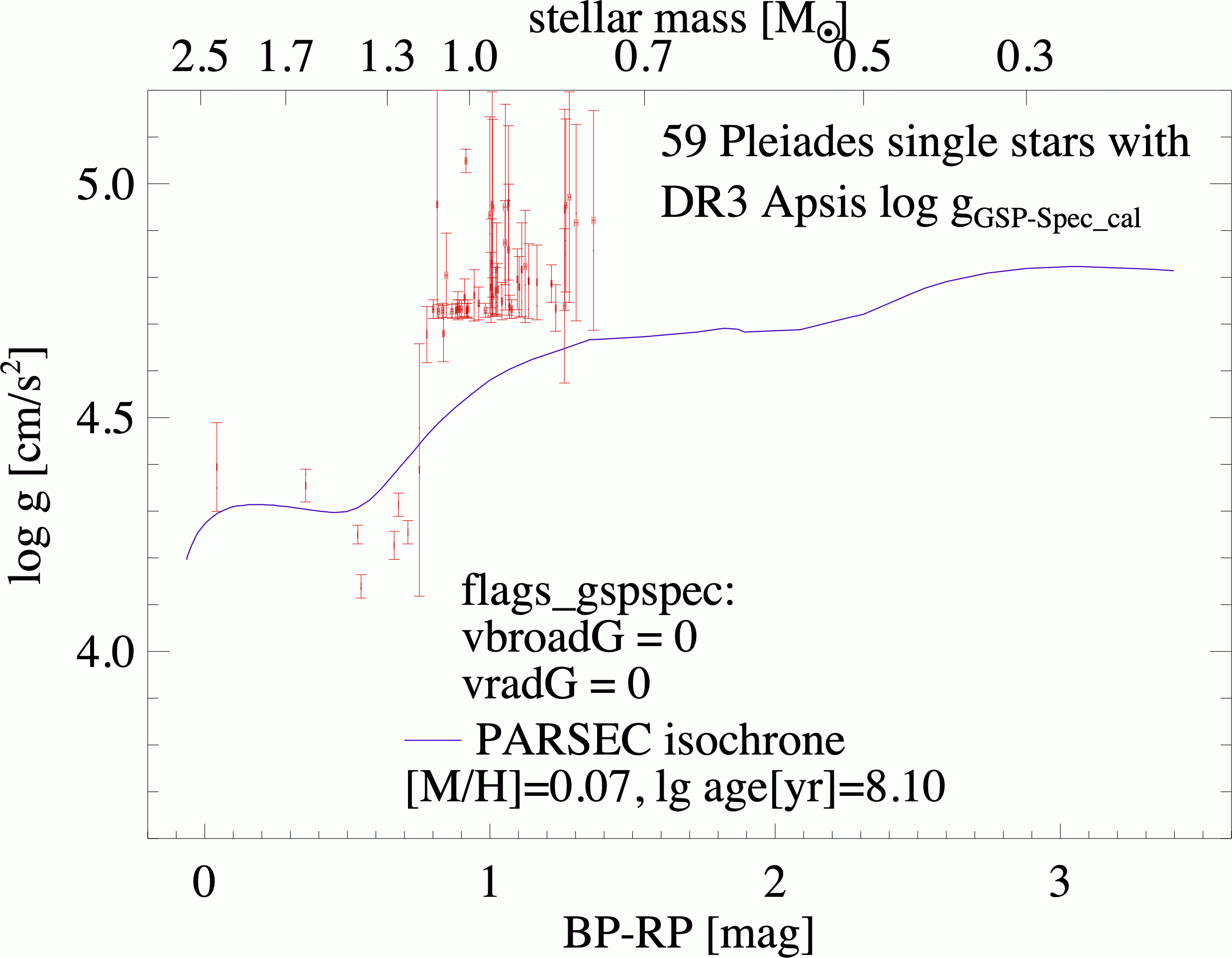}
}
}
\caption{Surface gravity estimates for single stars in the Hyades (top) and Pleiades (bottom) based on GSP-Phot (left) and recalibrated GSP-Spec (right). Not shown are stars with Apsis log\,g [cm/s$^2$] estimates below 3.8. Overplotted in blue is the best-fitting PARSEC isochrone.
\label{fig:Hyalogg}}
\end{figure*}

The top panel of Fig.\ \ref{fig:HyaTeff} shows the Apsis deduced effective temperatures for stars in the Hyades as a function of their BP-RP colour. On the left we show the GSP-Phot derived T$_{\rm eff}$, which reveals a monotonic and relatively smooth decline with increasing colour up to BP-RP$\lessapprox$1.35\,mag. In the range 1.35$<$BP-RP$<$3.0\,mag, the GSP-Phot temperature estimates bifurcate. Overplotted in blue is the T$_{\rm eff}$ versus\ colour relation according to the best-fitting PARSEC isochrone. Compared to the isochrone, GSP-Phot T$_{\rm eff}$ estimates are systematically lower by 150 to 250\,K for BP-RP$\lessapprox$1.35\,mag, while the bifurcated GSP-Phot T$_{\rm eff}$ estimates for BP-RP$>$1.35\,mag seemingly bracket the temperature according to the isochrone. For BP-RP$>$3.0\,mag, the GSP-Phot T$_{\rm eff}$ estimates are systematically higher by $\approx$50\,K than implied by the isochrone. On the right, we show GSP-Spec T$_{\rm eff}$, which also reveals a bifurcation for the colour range 1.35$<$BP-RP$<$1.65\,mag. 
For reference, the corresponding stellar masses according to the isochrone are indicated on the top abscissa.

The bottom panel of Fig.\ \ref{fig:HyaTeff} shows effective temperature versus\ colour for the Pleiades. There appears to be a noticeable gap in GSP-Phot derived T$_{\rm eff}$ in the range of 8500 to 9500\,K with the majority of the stars with BP-RP$<$0.3\,mag being assigned T$_{\rm eff}>$9500\,K. For redder colours, that is, 0.3$<$BP-RP$<$1.3\,mag,  GSP-Phot T$_{\rm eff}$ systematically falls a few hundred\,Kelvin  below the colour--temperature
relation of the best-fitting PARSEC isochrone, and for BP-RP$>$1.7\,mag it systematically
falls a few hundred Kelvin  above this relation. We note that a few stars in the BP-RP colour range from 1.0 to 2.0\,mag have a  GSP-Phot temperature that falls significantly above the isochrone relation. The GSP-Spec module was only able to deduce effective temperature for 115 of the bona fide single stars in the Pleiades. While the majority  closely follow the T$_{\rm eff}$--colour relation of the best-fitting isochrone, a few stars stand out by having significantly lower GSP-Spec temperatures.

The top panel of Fig.\ \ref{fig:Hyalogg} shows the surface gravity for stars in the Hyades as a function of their BP-RP colour as deduced by Apsis, and the left panel shows the log\,g  derived by  GSP-Phot. Similar to the T$_{\rm eff}$ estimates, log\,g follows a smooth, monotonic relation with colour for BP-RP$\lessapprox$1.35\,mag. In the colour range 1.35$<$BP-RP$<$3.0\,mag, GSP-Phot log\,g estimates bifurcate. For BP-RP$<$3.0\,mag, GSP-Phot log\,g\,[cm/s$^2$] estimates are typically 0.1 to 0.2 lower than suggested by the isochrone, while they fall above the isochrone value by $\approx$0.03 for redder colours. The right panel shows the recalibrated GSP-Spec log\,g estimates according to \citet{RecioBlanco2022}. On average, the recalibration lowers log\,g\,[cm/s$^2$] by $\approx$0.3 compared to the original DR3 Apsis estimates (e.g.\ from 5.0 to 4.7). The log\,g\,[cm/s$^2$] values stay almost constant at 4.7\, for 0.7$<$BP-RP$<$2.2\,mag, with the exception of a dip in the range of 1.35$<$BP-RP$<$1.70\,mag.

Shown in the bottom of Fig.\ \ref{fig:Hyalogg} are the corresponding plots for the Pleiades. GSP-Phot log\,g\,[cm/s$^2$]  estimates for BP-RP$<$1.7\,mag fall below the isochrone by up to 0.2. In the colour range 1.9$<$BP-RP$<$3.0\,mag, GSP-Phot log\,g estimates bifurcate. For red colours (BP-RP$>$3.0\,mag), the GSP-Phot log\,g estimates are systematically higher than suggested by the best-fitting isochrone. Akin to the single stars in the Hyades, the recalibrated GSP-Spec log\,g\,[cm/s$^2$] estimates for Pleiades members tend to cluster around 4.7.

\section{Discussion of benchmarking results} \label{sec:discussion}

\subsection{Age and FLAME}

The Apsis FLAME module is tuned using supervised learning \citep{Rybizki2020} based on templates, which include a model of the Milky Way. Stars get assigned a population identification (popID) derived from the age bins defined by \citet{Robin2003} for the different Galactic populations, namely\ thin disc (popID 0 to 6), thick disc (popID 7), halo (popID 8), plus additional popIDs for bulge stars and stars belonging to the Magellanic Clouds or to open clusters. The apparent age--colour relation in Fig.\ \ref{fig:HyaAge} suggests that the Apsis age inference is dominated by the model of the underlying Galactic population. FLAME seems to assign stars a popID, and hence an `age' primarily based on their effective temperature.

FLAME uses a solar [M/H] prior \citep{Creevey2022}, which should not be a major limitation to its applicability to the  near-solar [M/H] of the Hyades ([M/H] = 0.18$\pm$0.03) and Pleiades ([M/H] = 0.07$\pm$0.03) open clusters.
We note a potential mismatch between different stellar evolutionary models employed by GSP-Phot ---which provides inputs to FLAME--- and FLAME itself. According to \citet{Creevey2022}, GSP-Phot employs PARSEC 1.2S Colibri S37 models \citep{Tang2014,Chen2015,Pastorelli2020}, which is\ the same set of models we use for the determination of the stellar parameters of the single stars in the Hyades \citep{Brandner2023b} and Pleiades. However, FLAME itself employs Bag of Stellar Tracks and Isochrones \citep[BaSTI,][]{Hidalgo2018}.
Figure 12 in \citet{Hidalgo2018} highlights the close match of BaSTI and MIST isochrones \citep{Choi2016}, and the mismatch between BaSTI and PARSEC isochrones for young late-type stars of approximately solar abundance. 

As discussed by \citet{Brandner2023a}, MIST models tend to systematically predict fainter absolute magnitudes G$_{\rm abs}$ and bluer BP-RP colours for bona fide single stars in the Hyades with T$_{\rm eff} \le 4700$\,K compared to those observed by Gaia DR3. PARSEC isochrones provide a considerably better fit to the observed single-star sequence \citep{Brandner2023b}.

A further limitation of Gaia DR3 FLAME is the apparent lower age limit of 200\,Myr. This precludes a proper age dating of Pleiades members. In summary, individual FLAME age estimates seem of very limited informative power.

\subsection{Metallicity and GSP-Phot/GSP-Spec}

In addition to the low-resolution spectroscopic information in the BP and RP bands, GSP-Phot [M/H] estimates also incorporate Gaia parallax measurements, and use both theoretical model spectra and isochrones \citep{Andrae2022}. \citet{Fouesneau2022} caution against the use of GSP-Phot [M/H] estimates without additional investigation, and state that [M/H] estimates appear to be 0.2\,dex lower on average than literature data for sources with [M/H]$>-1$\,dex. This is in agreement with our findings. For both the Hyades and Pleiades samples, we find that the weighted mean [M/H] estimates of the bona fide single stars with GSP-Phot [M/H] estimates fall below the estimates derived from families of best-fitting isochrones by 0.15 to 0.2\,dex (Fig.\ \ref{fig:HyaMH}).

However, GSP-Phot [M/H] estimates of individual stars exhibit a much larger scatter than supported by the  16\% and 84\% quantiles of the Markov Chain Monte Carlo results reported in Apsis, suggesting that the uncertainties in the parameter estimates for individual stars are underestimated by a factor of between 10 and 20. A recalculation of [M/H] using the Python package gdr3apcal (Code Version 1.6) makes the discrepancy worse. We therefore agree with the recommendation by \citet{Fouesneau2022} to not use GSP-Phot [M/H] estimates for individual stars.

GSP-Spec [M/H] estimates are based on Gaia RVS and grids of model spectra spanning the full range of Galactic stellar populations. The inherent estimation bias of this module has been quantified by \citet{RecioBlanco2022}, who also provide a prescription for correction. The correction moves the ensemble [M/H] estimates of Apsis to within $\approx$0.1\,dex of the [M/H] values suggested from the isochrone fitting to the Hyades and Pleiades. For individual stars, in particular in the Hyades, GSP-Spec [M/H] (minimal) quantile uncertainties appear to be underestimated, which is an indication that the actual noise floor of the high-S/N estimates is considerably higher.

While the GSP-Phot and GSP-Spec estimates show a weak correlation  overall, individual estimates tend to disagree by substantially more than suggested by the assigned uncertainties. In general, neither GSP-Phot nor GSP-Spec is able to reproduce the [M/H] estimates derived from the isochrone fitting, which are marked by blue open squares. The apparent close match of the weighted mean $<{\rm [M/H]_{GSP-Phot}}>_{\rm wm} = 0.1720\pm 0.001$ for the Hyades is accidental, as it is the result of a selection of stars with both GSP-Phot and GSP-Spec estimates. A sample of 213 stars in the Hyades with GSP-Phot [M/H] estimates, and including 96 stars without GSP-Spec estimates, yields $<{\rm [M/H]_{GSP-Phot}}>_{\rm wm} = 0.034\pm 0.001$. Overall, both GSP-Phot and GSP-Spec tend to systematically underestimate [M/H] of the stars in the Hyades and Pleiades.

\subsection{Effective temperature and GSP-Phot/GSP-Spec}

As the spectral characteristics of stars are primarily determined by their effective temperature, it is no surprise that this is also the most robust of the DR3 Apsis parameter estimates. Nevertheless, GSP-Phot T$_{\rm eff}$ estimates seem to systematically fall by 150\,K to 250\,K below the T$_{\rm eff}$--colour relation suggested by the best-fitting isochrones for BP-RP$<$1.35\,mag. We find the apparent bifurcation of T$_{\rm eff}$  estimates for stars in the Hyades with  1.35\,mag$<$BP-RP$<$3.0\,mag and the apparent dearth of stars with 8000\,K$<$T$_{\rm eff} <$9500\,K in the Pleiades to be surprising. Apart from the systematic offsets, the GSP-Phot T$_{\rm eff}$ quantile uncertainties appear plausible (Fig.\ \ref{fig:HyaTeff}, left).
Overall, GSP-Spec estimates are in better agreement with the T$_{\rm eff}$--colour relation according to the PARSEC isochrone than GSP-Phot estimates, though estimates start to diverge for BP-RP$>$1.35\,mag (Fig.\ \ref{fig:HyaTeff}, right). 

\subsection{Surface gravity and GSP-Phot/GSP-Spec}

GSP-Phot log\,g estimates overall follow the log\,g--colour relation suggested by the best-fitting isochrones. We note the bifurcation in the log\,g estimates in the colour range of 2.0\,mag$<$BP-RP$<$3.0\,mag, and the systematic underestimation of log\,g for BP-RP$<$1.5\,mag (Fig.\ \ref{fig:Hyalogg}, left).

GSP-Spec log\,g\,[cm/s$^2$] estimates, which have been recalibrated according to the prescription provided by \citet{RecioBlanco2022} are almost constant at 4.7 over a wide colour range, with a strange dip to $\approx$4.4 at BP-RP$\approx$1.4\,mag for stars in the Hyades. We suspect that the lack of a dimension for rotational line broadening in the GSP-Spec model grid might force the parameter estimates for fast rotating stars in the Pleiades
in particular to higher apparent log\,g values (Fig.\ \ref{fig:Hyalogg}, right). The restriction to stars with GSP-Spec quality flags vbroadG and vradG equal to 0 (see \cite{RecioBlanco2022}) removes outliers in particular at the faint end. The systematics in the DR3 Apsis GSP-Spec estimates, though, are still present.

\subsection{Limitations of our method}
While the isochrone fits and the deduction of stellar astrophysical parameters from the isochrones provide robust estimates of stellar properties, there are also some inherent limitations to our approach. Firstly, the $\chi^2$ isochrone fitting is based on the assumption of uniform extinction to all stars in either of the clusters.\footnote{We note that due to their vicinity, the line-of-sight extinction is very small towards the Hyades (A$_{\rm V}$=3\,mmag), and modest for the Pleiades (A$_{\rm V}$=120\,mmag).} Apsis GSP-Phot, on the other hand, fits individual line-of-sight extinction estimates for each star. While this potentially provides fits of greater accuracy, it also opens an additional dimension for fitting errors due to a strong correlation between the uncertainties in T$_{\rm eff}$ and A$_{\rm V}$ estimates \citep{Andrae2022}. 

Secondly, our bona fide single star sequences are still contaminated by sources with hidden binarity. These are typically close binary systems with mass ratios and/or brightness ratios of $<$1:2. We expect the contamination to be more severe for the Pleiades sample due to its approximately three times larger distance, which reduces the sensitivity to binary detection and rejection based on the RUWE value. Still, the bias induced by the fainter and redder secondary on the parameter estimation in such systems should be modest, and cannot explain the systematic errors encountered in the Apsis estimates.

Thirdly, we employ non-rotating isochrones. Stellar rotation affects the upper main sequence and main sequence turn-off region in the Hyades and Pleiades  in particular, which are also the regions that are the most sensitive to the isochronal ages. However, for the majority of the stars in our sample, the effect on the determined astrophysical parameters should be minimal.

Finally, the PARSEC models follow a fixed Y(Z) abundance relation. However, this affects both our estimates and Apsis. We note that a deviation of the [Y/Z] abundance ratio from the relation would in particular change the energy-production rate of the low-mass stars \citep{Baraffe2018}.

\section{Recommendations and outlook} \label{sec:outlook}

\begin{table*}[htb]
\caption{Gaia DR3 Apsis quality assessment for main sequence stars with 11500\,K$\ge$T$_{\rm eff}\ge$3000\,K and [M/H]$\approx$0. \label{tab:Apsis_assess_tab}}
\centering 
\begin{tabular}{l l l}
Quantity & Typical offset & recommendation$^2$ \\ \hline
FLAME age & varies with spectral type, up to $>$10\,Gyr & do not use \\ \hline
GSP-Phot [M/H] & ensemble [M/H] 0.15 to 0.20\,dex too low & do not use for individual stars\\
GSP-Phot T$_{\rm eff}$  &$\approx$200\,K too low for BP-RP$<$1.35\,mag, bifurcates for BP-RP$>$1.35\,mag & ok with $\pm$250\,K uncertainty \\
GSP-Phot log\,g  & 0.1 to 0.2\,cm/s$^2$ too low for BP-RP$<$3.0\,mag & ok with $\pm$0.2\,cm/s$^2$ uncertainty\\ \hline
GSP-Spec [M/H]$^1$ &ensemble [M/H] $\approx$0.1\,dex too low  & ok with $\pm$0.2\,dex uncertainty\\
GSP-Spec T$_{\rm eff}$  &bifurcates for 1.35$<$BP-RP$<$1.65\,mag & ok  \\
GSP-Spec log\,g$^1$  & non-physical variation with spectral type & do not use\\
\end{tabular}
\tablefoot{$^1$Recalibrated according to \cite{RecioBlanco2022}. $^2$The uncertainties quoted in the column {\it recommendation} should be understood as systematic uncertainties in addition to the quantile estimates provided by Apsis. ``ok'' indicates that the estimates are largely in agreement with estimates according to PARSEC stellar models.}
\end{table*}

While Gaia offers unprecedented quality and precision in its astrometric, photometric, and spectroscopic measurements, we identify tension between Gaia DR3 Apsis estimates and astrophysical properties deduced by us for $\approx$1500 bona fide single stars in the Hyades and Pleiades open clusters. 

Apsis provides very good to fair estimates of the effective temperature of stars, and potentially of the ensemble [M/H] of homogeneous stellar groups. However, both GSP-Phot and GSP-Spec seem to underestimate the ensemble [M/H] of two nearby open clusters by 0.1 to 0.2\,dex. As recommended by \citet{Andrae2022}, one should not use the GSP-Phot [M/H] for individual stars.
GSP-Phot log\,g in general is sufficiently precise to deduce luminosity classes, while GSP-spec log\,g\,[cm/s$^2$] for stars in the open clusters converge around 4.7 over a larger range of spectral types.  Our assessment of Gaia DR3 Apsis and the resulting recommendations are summarised in Table \ref{tab:Apsis_assess_tab}. 

We also identified a few potential issues in the underlying assumptions and training sets used for Gaia DR3 Apsis.
Firstly, GSP-Phot employs PARSEC isochrones. FLAME relies on BaSTI models, whose isochrones closely resemble MESA isochrones but disagree with PARSEC isochrones for solar metallicity stars with masses of $\lesssim$0.7\,M$_\odot$. This model mismatch might bias FLAME estimates. Secondly,
the GSP-Spec spectral model grid lacks a dimension for rotational broadening, and therefore fast-rotating stars might be assigned excessive log\,g estimates.
Finally, FLAME age estimates reproduce the training model. The ages start at 100\,Myr according to the online documentation, while the youngest (16th quantile) ages assigned to Pleiades members are 200\,Myr. FLAME lacks the ability to classify the youngest (age $\lesssim$150\,Myr) Galactic population (popID = 0 according to \citet{Rybizki2020}).

Addressing these issues might help to improve the quality of Apsis estimates. Overall, DR3 Apsis seems to assign overly small uncertainties to its estimates. Therefore, Gaia DR3 Apsis cannot replace more dedicated efforts to determine stellar properties \citep[e.g.][]{Anders2022,Berger2023}, or for example provide the level of accuracy required for the characterisation of exoplanet host stars.

We expect that the sample of 1500 bona fide single stars in the Hyades \citep{Brandner2023b} and Pleiades open clusters with homogeneous astrophysical parameter estimates derived from PARSEC evolutionary models might also be highly useful for future benchmarks of theoretical and observational studies.

\begin{acknowledgements}
This work has made use of data from the European Space Agency (ESA) mission {\it Gaia} (\url{https://www.cosmos.esa.int/gaia}), processed by the {\it Gaia} Data Processing and Analysis Consortium (DPAC, \url{https://www.cosmos.esa.int/web/gaia/dpac/consortium}). Funding for the DPAC has been provided by national institutions, in particular the institutions participating in the {\it Gaia} Multilateral Agreement. We acknowledge the use of TOPCAT \citep{Taylor2005,Taylor2011}.
\end{acknowledgements}

\bibliographystyle{aa}
\bibliography{lit}

\end{document}